# Volumetric and Simultaneous Photoacoustic and Ultrasound Imaging with a Conventional Linear Array in a Multiview Scanning Scheme

Clément Linger, Yoann Atlas, Remy Winter, Marine Vandebrouck, Maxime Faure, Théotim Lucas, S. Lori Bridal, and Jérôme Gateau[1]

*Abstract*— **Volumetric, multimodal imaging with precise spatial and temporal co-registration can provide valuable and complementary information for diagnosis and monitoring. Considerable research has sought to combine 3D photoacoustic (PA) and ultrasound (US) imaging in clinically translatable configurations. However, technical compromises currently result in poor image quality either for photoacoustic or ultrasonic modes. This work aims to provide translatable, high quality, simultaneously co-registered dual-mode PA/US 3D tomography. Volumetric imaging based on a synthetic aperture approach was implemented by interlacing PA and US acquisitions during a rotate-translate scan with a 5-MHz linear array (12 angles and 30-mm translation to image a 21-mm diameter, 19 mm long cylindrical volume within 21 seconds). For co-registration, an original calibration method using a specifically designed thread phantom was developed to estimate 6 geometrical parameters and 1 temporal off-set through global optimization of the reconstructed sharpness and superposition of calibration phantom structures. Phantom design and cost function metrics were selected based on analysis of a numerical phantom, and resulted in a high estimation accuracy for the 7 parameters. Experimental estimations validated the calibration repeatability. The estimated parameters were used for bimodal reconstruction of additional phantoms with either identical or distinct spatial distributions of US and PA contrasts. Superposition distance of the two modes was within < 10% of the acoustic wavelength and a wavelength-order uniform spatial resolution was obtained. This dual-mode PA/US tomography should contribute to more sensitive and robust detection and follow-up of biological changes or the monitoring slower-kinetic phenomena in living systems such as the accumulation of nanoagents.**

## I. Introduction

**V**OLUMETRIC and simultaneously co-registered multimodal imaging is increasingly developing in biomedical imaging to more fully exploit the growing range of rich, multiplexed and complementary anatomical–functional information that can be precisely correlated in time and with respect to spatial position [1]. One of the first such bimodal imaging systems arrived in the clinical armamentarium in the 1990s. This pioneering multimodal approach, positron emission tomography (PET) coupled with computed tomography (CT), showed the interest of simultaneous co-registered tomography providing the anatomical context of CT to better interpret metabolic information offered by PET [2]. Many other original volumetric multimodal tomography combinations have followed. For example, PET has been integrated with magnetic resonance imaging (MRI) [1], [2]. Photoacoustic imaging (PAI) and optical coherence tomography (OCT) have been superimposed [3] to enable novel extraction of tissue characteristics like chromophore concentration. PET has been combined with Doppler ultrasound imaging [4] to relate metabolism and blood flow.

Multimodal tomography that provides a simultaneous and co-registered view of different aspects within the body can provide manifold advantages. By obtaining more information within a single imaging session, patients, researchers and clinical-management teams benefit due to both reduced examination times and more comprehensive characterization of the examined region. Precise co-registration in space and time is essential for a characterization that makes the most of each modality. In addition, volumetric imaging provides a detailed view of regions under examination from various orientations for improved diagnosis and facilitates comparisons during longitudinal studies or monitoring of therapeutic response.

Ultrasound (US) and PA imaging modalities display complementary information such that when combined, US imaging delineates organs and lesions to provide the anatomical reference frame for the PA-based information which can range from the characterization of hemoglobin oxygenation to the detection of molecular and nanoparticular contrast agents [5], [6]. Although the recorded ultrasonic signals are generated *in situ* by the optical absorption of a laser excitation for PA imaging, while, for US, they are created by transmitting an ultrasonic pulse that is then backscattered from structures with different acoustic impedances, both modalities rely on the detection of ultrasonic signals. This is an advantage for bimodal PA/US imaging because signals can be received using the same ultrasound detector for simultaneous co-registration. Such simultaneous co-registration has previously been demonstrated in 2D using a single US detector array [7], [8] and using a dual array configuration [9].

[T]This work was supported in part by Sorbonne University under the program Emergence Sorbonne Université 2019–2020. This project has received financial support from the CNRS through the MITI interdisciplinary programs (Defi Imag'IN, 80 Prime) and from Gefluc Paris, Ile de France This work was partly funded by France Life Imaging under Grant ANR-11-INBS-0006.

Bimodal PA/US in 2D, however, is hampered due to several limitations. Firstly, out-of-plane artifacts can be stronger in 2D PA imaging than in 2D US imaging. The reason is two-fold. Because light is strongly scattered within biological tissue (photon diffusion regime [10]), the optical excitation used in PA imaging is intrinsically 3D in nature. Furthermore, the ultrasound focusing in the elevation direction is weaker for PA imaging because it only occurs in reception mode. Secondly, elongated structures like blood vessels have a strong directionality in PA imaging and so they may not be visible in the limited-view configuration offered by most 2D PA imaging systems if the vessels emit outside of the limited angular aperture of the detector.

Spherical US detector matrices (2D arrays) provide a large angular aperture and have been developed specifically for 3D PA imaging using transducers that can also be used in pulse-echo mode to create US images [11], [12]. Given the limited number of elements used to cover the spherical surface, spherical US matrices can be considered sparse for ultrasonic imaging. However, pulse-echo US images of biological tissue are densely filled with signals from echogenic structures such that high-quality, 3D US imaging requires detector matrices with a higher spatial sampling than the spherical matrices designed for PA imaging. Thus, although a large, sparse angular aperture can provide high quality photoacoustic images, especially for angiographic applications [11], [13], [14] and, although such spherical arrays have been shown to perform well when used to extract US Doppler signals [15], [16] or the signals from sparsely distributed US contrast agents [15], their use for biological tissue anatomical imaging can result in strong side lobes and grating lobes artifacts and poor contrast.

Planar US transducer matrices can offer a less sparse transducer distribution. They have been developed for 3D US imaging and have been tested for 3D PA imaging [17], [18]. While US transducer matrices enable high-frame rate 3D US imaging and associated imaging modes (shear wave elastography, Doppler, localization microscopy …) [19], the limited angular aperture of such matrices and the poor sensitivity of the relatively small elements lead to limited-view artifacts, limited spatial resolution and poor sensitivity for 3D PA imaging.

Thus, at the present time, simultaneous co-registration of volumetric PA and US imaging with either spherical or planar US matrices results in degraded image quality for one of the two modalities. Specific mechanical scanning patterns with US linear arrays that have been designed for 2D imaging may provide an alternative solution toward high-quality imaging with both PA and US techniques. Several systems based on translating a linear array transducer along the elevational direction (perpendicularly to the imaged plane) have been developed recently and their ability to provide simultaneously co-registered 3D US and PA imaging has been demonstrated [20]–[23]. The 2D images are then stacked to obtain a volume. This approach is relatively easy to implement and to transfer to a clinical environment. However, the angular aperture in the elevational direction of a linear US array is very limited because of the weak elevational focus implemented to improve the sensitivity and to limit the elevational thickness of the imaged slice along a large range of depths. In PA imaging, directional structures could emit outside of this limited angular aperture which can cause strong limited-view artifacts. This issue is not addressed with the translational scan, even when multiple scan positions are used to reconstruct each slice (synthetic aperture focusing) [24], [25], because the orientation of the angular aperture remains constant for all the scan positions. Moreover, the spatial resolution in the translational direction is strongly degraded compared to the in-plane resolution both for PA and US imaging.

Adding a rotational motion and implementing a rotate-translate synthetic aperture scanning of a linear US array has been shown to effectively increase the angular aperture and to highly improve the volumetric image quality and elevational resolution compared to a purely translational scan [26], [27]. Synthetic aperture scanning refers here to the fact that signals acquired at several scan positions are coherently combined during the reconstruction process to synthetize a larger angular aperture. The rotate-translate approach has previously been demonstrated independently on two different imaging systems for 3D PAI [26] and for 3D USI [27], but the data acquired with these systems has not yet been co-registered to provide dual-mode imaging.

In this paper, we propose and demonstrate a dual-mode PA and US imaging approach with a linear US array that provides simultaneously co-registered volumetric data with high quality for both modes throughout a volume on the order of 7 cm$^3$ and within 21 s. A thread phantom calibration identifies the 7 required reconstruction parameters based on cost-function global optimization. We present here the first dual PA-US 3D calibration method. Calibration is conceived to be durable as long as the transducer is not repositioned in its support, even if the optical source is realigned. Since this dual-mode PA and US imaging approach is based on a single rotate-translate scan of a linear US array and does not call upon fiducial points for co-registration, this technique could be readily adapted to *in vivo* imaging constraints for a variety of applications.

## II. MATERIALS AND METHODS

### A. *Experimental set-up*

The experimental setup presented in Fig. 1 (a) consists of : (1) the optical excitation system comprised of a nanosecond laser and an optical fiber bundle, (2) the US excitation and acquisition system composed of an US linear array connected to a programmable US platform, (3) the scanning system made up of two motorized stages operated by a motion controller, and (4) the synchronization based on a programmable trigger generator. The entire acquisition process was automated.

An optical excitation at 700 nm was generated using an optical parametric oscillator laser (SpitLight 600 OPO, Innolas Laser GmbH, Krailling, Germany) delivering < 8 ns pulses with a pulse repetition frequency (PRF) of 20 Hz. A bifurcated fiber bundle (CeramOptec GmbH, Bonn, Germany) guided the light toward the imaged volume to provide bilateral illumination.

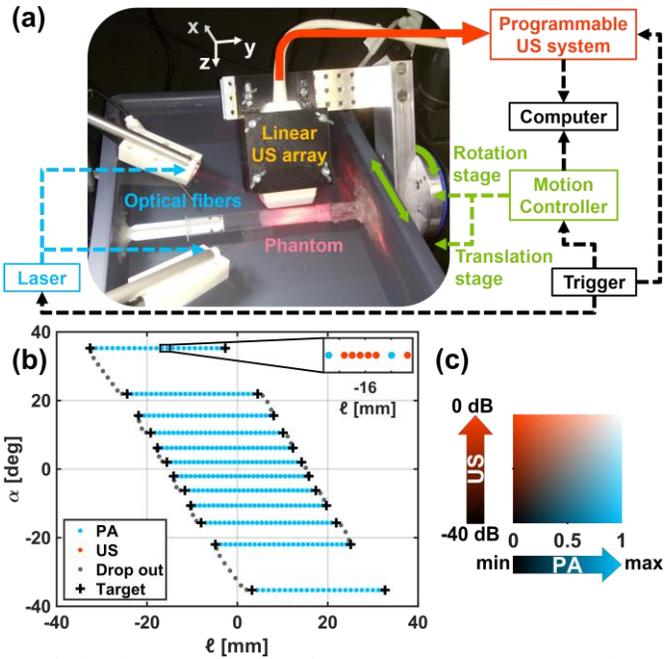

Fig. 1. (a) Annotated picture of the experimental setup. (b) Motor positions for the PA acquisitions over one scan. For a better readability, positions corresponding to the US and PA acquisitions are indicated only for the inset in the upper right corner of the graph. Positions shown in grey (drop out) were not used for the reconstruction, but are acquired due to the continuous motion of the motors (c) Colormap used for all images: PA images are represented in shades of blue and US images in shades of orange, the sum of the two leading to white. PA signals are presented on a linear scale while US signals are presented within the range from -40 to 0 dB.

The mean laser energy at each fiber output was estimated to be around 6 mJ. The fixed PRF of the Laser sets the time base for the acquisition sequence. The pulse energy was recorded using a pyrometer incorporated in the laser.

A 128-element linear US transducer array (L7–4, ATL, Seattle, WA, USA) was driven by a programmable, 64-channel US machine (Vantage, Verasonics, WA, USA). For all the transmit events and all the receive events, only the 64, central elements of the array were used. Each Laser pulse triggered a receive-only event to record the PA data. The 50-ms interval between two laser pulses, was divided into equal 6.25-ms parts for eight events: the initial PA acquisition, then a 6.25-ms pause followed by 5 US "plane wave" (a beam that is unfocused in the lateral direction of the array) transmit-receive events at a PRF of 160 Hz [28] and at steered angles of -4°, -2°, 0° (the 64 elements in the transducer array were fired at the same time), 2° and 4°, respectively, a second 6.25 ms pause and finally a new PA acquisition. The pauses just before and after laser pulses prevented possible interference between the ultrasound fields generated by the laser pulse (PA) and by the plane wave emissions (US). Since the array is continuously scanned, the low PRF for the US acquisitions enables further diversification of the spatial positions [27]. Transmitted US pulses were 1 cycle long with a center frequency of 5.2 MHz. Pulse-echo US signals and PA signals were recorded at a sampling frequency of 20 MS/s and 62.5 MS/s, respectively. The gain was adjusted to a constant value so that all received PA and US signals had a sufficient amplitude for good digitalization without risk of saturation. No time gain compensation was applied because of the weak ultrasound attenuation of the imaged samples.

The rotate-translate scanning system described previously in ref [27] consisted of a US array fixed to a rotation stage connected to a translation stage (Physik instrumente, Karlsruhe, Germany), so that the rotation axis was perpendicular to the translation axis. The array axis (axis along the row of its elements) was aligned with the rotation axis. The rotation stage was moved to 12 different angles with a nominal angular sampling step of $\Delta\alpha = 4°$. At each angular position, the imaged volume was scanned with a translation range L = 30 mm and a linear sampling step $\Delta\ell = 1$ mm. Relative to parameters used in ref [27], the scan parameters, in particular the translation velocity, had to be adjusted to adapt for the PRF of the Laser. The motion was continuous, to provide more time-efficient scanning compared to stepped motion [29]. For each event producing US signals (laser emission and series of 5 US plane wave emissions), the motor positions were recorded and stored in the motor controller to be used for the image reconstruction. The scan was automated and initiated with an external trigger sent simultaneously by a generator (BNC Model 577, Berkeley Nucleonics, San Rafael, CA, USA) to the Laser, the programmable US system and the motion controller to synchronize all devices and the emission-acquisition sequence. The motor positions are presented graphically in Fig. 1(b).

The array elements were acoustically coupled to the imaged sample in a tank filled with tap water. The tips of the fiber bundles were also under water to provide close illumination of the sample. The water temperature was monitored based on readings from a thermometer (HI98509, Hanna instruments, Lingolsheim, France).to estimate the speed of sound in the water path [30].

To scan a full volume, 411 laser/ultrasound transmit sequences were fired in 21 s. After scanning, radiofrequency signals received after PA and US events, corresponding motor positions and pyrometer values were recorded for subsequent processing and image reconstruction. Video 1 is provided to show the movement of the transducer and the flashes for PA imaging.

*B. Image reconstruction*

*1) Image Reconstruction Algorithm*

The 3D image grid was defined in a fixed Cartesian coordinate system (O, $e_x$, $e_y$, $e_z$). The vectors $e_y$ and $e_z$ correspond to the rotation axis and the radial direction, respectively, when the rotation angle equals zero. The vector $e_x$ completes the orthonormal basis. The origin O is chosen as the orthogonal projection of the center of the array on the rotation axis, when the translation stage is at its central position ($\ell=0$). In this grid, the voxel dimensions ($p_x \times p_y \times p_z$) were selected to be 71 µm × 143 µm × 71 µm to correspond to the expected, anisotropic resolution along each axis. The image volume is defined by a diamond-shaped, cross-sectional area (DSCA, Fig. 2 (c)) in the xz-plane (length of the diagonal L= 30 mm, and centered at (x=0, z =25 mm)) and the active length of the array (19 mm) along the y-axis.

PA signals were divided by the corresponding pyrometer value to compensate for the pulse-to-pulse energy fluctuations of the Laser and bandpass filtered between 2 MHz and 10 MHz (Butterworth, order 3).

Image reconstruction was performed with delay-and-sum beamforming algorithms. The one-way (PA) and two-way (US) travel times between the US transducer element positions ($x_n$, $y_n$, $z_n$) and each imaged voxel ($x_p$, $y_p$, $z_p$) were computed, assuming a constant speed of sound, c, in the medium. Each voxel's value resulted from the sum of signals received by all array elements and tomographic positions arriving within the time-of-flight range estimated to correspond to propagation times to the voxel. The US image reconstruction algorithm was detailed in ref [27] and assumes that the "plane wave" emissions by the linear array correspond to cylindrical waves in the 3D space. For both PA and US image reconstruction, the apodization included a dynamic aperture along the axis of the array to maintain a constant angular aperture (lateral f-number of 1.3 with a Hamming window) and a mask to account for the elevational focus of the array (elevational thickness of 1.2 mm with a 20% cosine taper).

Three-dimensional, envelope-detected images were obtained and were displayed using maximum amplitude projection (MAP) along the coordinate system axes. Rotating MAP images around the z axis were obtained with the 3D project option of ImageJ [31]. The colorscale used for the images is displayed in Fig. 1(c).

### 2) Time delays for the reconstruction

The recording of the ultrasound signals was set to start at the same time as the Laser emission for the PA acquisition or at the same time as the ultrasound emission for the US acquisition. However, due to the lens effects [32] and additional time delays induced by the acquisition hardware, we found that two effective temporal offsets, $t_{0\,PA}$ and $t_{0\,US}$, needed to be determined to convert the voxel-associated travel times into the time indexes of the recorded signals. The offset, $t_{0\,PA}$, was determined experimentally. Two 20-µm diameter black nylon threads (NYL02DS, Vetsuture, France) were positioned perpendicularly to the imaging plane of the array and in the vicinity of the elevation focus (around 25 mm from the face of the array). They were illuminated with the Laser light and PA signals were recorded. Given their small diameter with regards to the wavelength at the center frequency of the array ($\Lambda_c \approx 300$µm), the threads can be assumed to be point absorbers in 2D. 2D images ($p_y \times p_z = 71$ µm × 71 µm) were beamformed for different values of $t_{0\,PA}$, using the temperature-adjusted speed of sound in the water bath. The one-dimension Brenner's gradient of the images was computed along the lateral dimension of the array:

$$F_{Brenner\,1D} = \sum_{y,z}(f(y+2,z) - f(y,z))^2 \quad (1)$$

where $f(y,z)$ represents the gray level intensity in the 2D image. The Brenner's gradient provides a quantitative measurement of image sharpness and has been shown to be an efficient metric for the speed of sound calibration in PA [33]. It was maximized for $t_{0\,PA} = -1.3$ µs. In the recorded signals, $|t_{0\,PA}|$ corresponded also to the time of arrival of PA signals generated by a metalized mylar film (space blanket) pressed against the face of the array.

### 3) Spatial transformation from motor to array positions

For each tomographic position, the experimental acquisition system records the motor positions in their own coordinate systems (integrated position sensors). However, the reconstruction algorithm requires the position ($x_n$, $y_n$, $z_n$) and orientation of the array elements within the fixed coordinate system (O, $e_x$, $e_y$, $e_z$) describing the image grid.

Each tomographic position is described by the linear and angular motor positions ($\ell$, $\alpha$), respectively. The translation length equals zero ($\ell=0$) when the translation stage is at the center position, and the rotation angle equal zero ($\alpha=0$) when the radial axis is vertical. The rotation axis is by definition parallel to $e_y$. We name $\mathcal{R}_\alpha$ the rotation matrix around the y-axis due to the motion of the rotation motor. The unit translation vector **t** is expected to be $e_x$ but mechanical misalignments requires the addition of two angular parameters: φ, the (azimuthal) angle measured from $e_x$ to the orthogonal projection $t_p$ of **t** on the xz-plane, and θ, the angle from $t_p$ to **t**. Therefore,

$$\begin{pmatrix} t_x \\ t_y \\ t_z \end{pmatrix} = \begin{pmatrix} \cos(\theta) \cdot \cos(\varphi) \\ \sin(\theta) \\ \cos(\theta) \cdot \sin(\varphi) \end{pmatrix} \quad (2)$$

The central point of the active transducer aperture is named $O_a$, and is located on the interface between the transducer array and the water. The coordinates of $O_a$ in (O, $e_x$, $e_y$, $e_z$) depend on ($\ell$, $\alpha$). We define $\Delta x = x_{Oa}(\ell=0, \alpha=0)$ and $\Delta z = z_{Oa}(\ell=0, \alpha=0)$.

$$\begin{pmatrix} x_{Oa}(\ell,\alpha) \\ y_{Oa}(\ell,\alpha) \\ z_{Oa}(\ell,\alpha) \end{pmatrix} = \mathcal{R}_\alpha \cdot \begin{pmatrix} \Delta x \\ 0 \\ \Delta z \end{pmatrix} + \ell \cdot \begin{pmatrix} t_x \\ t_y \\ t_z \end{pmatrix} \quad (3)$$

Additionally, the spatial transformation matrix from the mobile Cartesian coordinate system ($O_a$, **u**, **v**, **w**) attached to the transducer array to the fixed coordinate system (O, $e_x$, $e_y$, $e_z$) needs to be assessed. The vector **u** corresponds to the elevation direction, the vector **v** to the long axis along the row of the elements and the vector **w** to the axial direction of the linear array. The transformation matrix is written as:

$$\begin{pmatrix} u_x(\alpha) & v_x(\alpha) & w_x(\alpha) \\ u_y(\alpha) & v_y(\alpha) & w_y(\alpha) \\ u_z(\alpha) & v_z(\alpha) & w_z(\alpha) \end{pmatrix} = \mathcal{R}_\alpha \cdot \mathcal{R}_{Pitch} \cdot \mathcal{R}_{Yaw} \cdot \mathcal{R}_{Roll} \quad (4)$$

$\mathcal{R}_{Roll}$, $\mathcal{R}_{Pitch}$ and $\mathcal{R}_{Yaw}$ are rotation matrices around the 1st-axis, the 2nd-axis and the 3rd-axis, respectively. The product of these rotation matrices models the misalignment of the array axis **v** compared to the rotation axis of the stage $e_y$.

Therefore, the position of the center of the element number *n* of the array ($n \in [\![1,N]\!]$ with N=64) in (O, $e_x$, $e_y$, $e_z$) can be decomposed as:

$$\begin{pmatrix} x_n \\ y_n \\ z_n \end{pmatrix} = \begin{pmatrix} x_{Oa}(\ell,\alpha) \\ y_{Oa}(\ell,\alpha) \\ z_{Oa}(\ell,\alpha) \end{pmatrix} + \left(n - 1 - \frac{N-1}{2}\right) \cdot p \cdot \begin{pmatrix} v_x(\alpha) \\ v_y(\alpha) \\ v_z(\alpha) \end{pmatrix} \quad (5)$$

With *p* the interelement spacing of the array. Here, $p = 298$ µm.

Finally, a total of 7 geometrical parameters independent of ($\ell$, $\alpha$) need to be determined: Roll, Pitch, Yaw, $\Delta x$, $\Delta z$, θ, φ.

### C. Calibration method

Inadequate estimation of reconstruction parameters results in degraded image quality in terms of sharpness and will lead to misalignment of the PA and US images. Therefore, we developed a calibration method based on a specific phantom and an optimization algorithm to determine the required 7 reconstruction parameters: $t_{0\,US}$, Roll, Pitch, Yaw, $\Delta x$, $\Delta z$, θ. The parameter φ was set to be equal to zero because the perpendicularity between the translation axis and $\mathbf{e_z}$ was precisely ensured by the mechanical design, and because solutions for our optimization algorithm are equivalent for different sets of (Pitch, φ) values.

The calibration method only needs to be repeated if the position of the array within the scanning system is readjusted (for example by removing the transducer from the support). The calibration remains valid, however, if the position of the illumination is modified relative to the ultrasound array.

### 1) Calibration Phantom

The calibration phantom needs to be simple to build, easy to use and should not require an absolute and tedious positioning procedure. We, therefore, developed a wire phantom, inspired by phantoms used for the calibration of freehand 3D ultrasound systems such as Z-fiducial phantoms [34]. Wires or threads have several advantages. First, black threads have strong contrast compared to water for both US and PA imaging, and are therefore expected to be clearly visible for both modes and superimposed on the dual-modality images. Second, straight threads provide elongated and uniform structures that can be easily intersected and identified in a volumetric image even with a sparse sampling in one direction (as opposed to small spheres for instance). Third, their orientation can be varied. Finally, by using a few non-crossing and well-separated threads, the segmentation of the 3D image allows local assessments of the image quality by characterizing the image sharpness in specific zones that intersect a thread.

Our calibration phantom is presented in Fig. 2. It is comprised of four threads with two orientations: two threads parallel to one another and positioned in a horizontal plane (Thread 1 and 3), and two threads rotated by ± 22° placed in parallel planes below and above, respectively (Thread 2 and 4). The angle was chosen to provide good sensitivity to the different parameters to be estimated. The spacing between the threads was chosen so that the phantom fits inside the imaged volume when the parallel threads are roughly aligned along the y-axis and at z ≈ 25 mm. The orientation of the phantom was chosen so that the threads appear as point-like shapes in xz-planes whose reconstruction sharpness is highly sensitive to the tomographic positions.

The experimental calibration phantom was implemented with 20-µm diameter black nylon threads (NYL02DS, Vetsuture, France) mounted on a 3D-printed frame (Fig. 2 (b)).

### 2) Numerical Calibration Phantom

Because the values of the parameters estimated experimentally with the calibration phantom could not be determined with a reference measurement, we validated the calibration method using a numerical simulation in which the set of parameters $\chi_c = (t_{0\ US\ c}, Roll_c, Pitch_c, Yaw_c, \Delta x_c, \Delta z_c, \theta_c)$ is an input.

For the input set of parameters $\chi_c$, $\varphi_c = 0$ and a given motor position $(\ell, \alpha)$, the orientation $(\mathbf{u}, \mathbf{v}, \mathbf{w})$ of the array and the position $(x_n, y_n, z_n)$ of each element were computed from equations (2)-(5). An experimental dataset of motor positions was used to simulate the entire scan.

The simulation of the ultrasound signals (PA or US) for a given motor position and a given element of the array was based on a semianalytical method [35] for linear wave propagation in a 3D homogeneous medium without

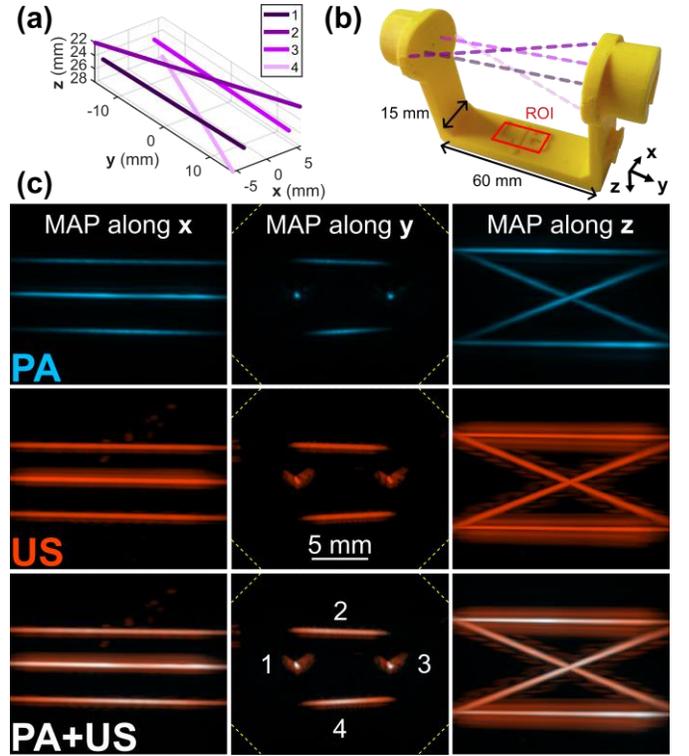

Fig. 2. (a) Imaged region (also called region of interest (ROI)) of the calibration phantom in a 3D coordinate system with dimensions in mm. (b) Picture of the calibration phantom. Four 20-µm nylon threads are mounted on a yellow 3D-printed frame. To ease the readability, the threads have been highlighted with the same color as in (a). A projection of the ROI is shown in red. (c) Volumetric images reconstructed with a set of optimized parameters: first row: photoacoustic image; second row: ultrasound image; third row: Combined PA/US image. The colorscale is presented in Fig. 1 (c). Each image is a maximum amplitude projection (MAP) image. The visible part of the diamond-shaped, cross-sectional area (DSCA) is shown with yellow dashed lines in the MAP images along y. First column images are 19-mm wide (y-axis) by 16-mm high (z-axis) and centered at (y=0 mm, z=25 mm); second column images are 19- wide (x-axis) by 16-mm high (z-axis) and centered at (x=0 mm, z=25 mm); third column images are 19-mm wide (y-axis) by 16-mm high (x-axis) and centered at (x=0 mm, y=0 mm).

attenuation. To limit the frequency content of the simulated signals, we calculated the convolution of the impulse response directly for a specific waveform: a one cycle sinusoidal signal at 5 MHz with a gaussian envelope for the US simulation, and the derivative of a gaussian pulse (standard deviation 17 ns) for the PA simulation. Simulated data were generated at the same sampling frequency as the experimental data

The 3D spatial impulse response of the transducer was computed based on the discretization of the Rayleigh integral over the surface of the transducer. The finite size (250 µm-width and 7.5 mm-height) and cylindrical focusing of one array element (focal length 25 mm) were modeled for a cylindrical surface discretized with point transducers spaced by $\Lambda_c/5$. $\Lambda_c$ is the ultrasound wavelength at the center frequency of the transducer ($\Lambda_c \approx 300$ µm). The surface was oriented according to the basis $(\mathbf{u}, \mathbf{v}, \mathbf{w})$ and its center was positioned in $(x_n, y_n, z_n)$. We computed the impulse response of this array element for a given point $(x_s, y_s, z_s)$ in the imaged volume by summing the impulse responses over all the point transducers. For the PA simulation, the point $(x_s, y_s, z_s)$ was considered as a PA source. For the US simulation, pulse-echo signals of $(x_s, y_s, z_s)$ were computed for a tilted "plane wave" emission. The "plane wave" generation was computed by summing spatial impulse responses (delayed to induce the steered angle) over all the array elements. The pulse-echo signal for the element number $n$ was then obtained by convolving this sum with the spatial impulse response of $n$.

To simulate the acquisition of the PA and US signals for the calibration phantom, the threads were discretized along their length into point sources (PA) or point scatterers (US) spaced by $\Lambda_c/5$. The diameter of the threads was not modeled. The superposition principle was used to compute the signal for one element of the array at a given position: the simulated signals were added over all the points ($x_s$, $y_s$, $z_s$) corresponding the threads.

The input of the numerical simulation is: $\chi_s$, the motor positions ($\ell$, $\alpha$) and their corresponding events (PA acquisition or US acquisition with a tilted "plane wave" emission at a given steering angle), the speed of sound and the positions of the point sources (PA) or point scatterers (US). The output is the ultrasound signals corresponding to all the PA and US events received by each element of the array. Simulated signals were reconstructed using the method presented in section II.B.1. Point-like objects and threads were found to be reconstructed at their defined positions both in PA and US images, which validated the simulation method.

The numerical calibration phantom reproduced the spatial arrangement of the threads in the experimental calibration phantom (Fig 2a).

### 3) Calibration algorithm: Cost function

The calibration algorithm is based on an optimization algorithm that minimizes a cost function. The cost function was assessed using reconstructed volumetric images from one complete data acquisition sequence, comprised of PA and US (5 steered angles) data sets for the calibration phantom.

#### a) Calibration images and segmentation of the threads

Each evaluation of the cost function for a set $\chi$ of the 7 parameters required volumetric image reconstruction. The time needed for this was reduced using sparse sampling of the imaged volume along the y-axis. We reconstructed only 5 slices located at $y = i \times 3$ mm with $i \in [\![-2;2]\!]$, i.e. centered and distributed to avoid edge effects. For each slice, we reconstructed two 30-mm width square images (one PA and one US) centered around ($x = 0$, $z = 25$ mm) and with pixel sizes py × pz = 71 μm × 71 μm using the whole dataset and the 3D reconstruction algorithm described in section II.B.1.

Threads were segmented and identified to calculate several image metrics. Each envelope-detected image was thresholded at one fourth of its maximum pixel value to produce a binary image. The fourth largest connected components of the binary image were identified as regions and each region was expected to correspond to one thread. The cost was directly set to zero, meaning that the set $\chi$ is rejected, if less than four connected components were counted. For each region, the centroid position $(x_j, z_j)$ was calculated with weights based on the grayscale image intensity value. The distance between the centroids determined in the US image and in the PA image in slice $i$ was:

$$d_{\text{US-PA}}(i,j) = \sqrt{\left(x_j^{US} - x_j^{PA}\right)^2 + \left(z_j^{US} - z_j^{PA}\right)^2} \quad (6)$$

In each image, the four centroids were sorted by increasing angular position and each was attributed to one of the threads.

#### b) Selection process for image-based metrics with the numerical calibration phantom

Since no standard cost function exists for this 7-parameter optimization problem, the cost function was built empirically from observations of how some variable metrics, computed from the images, depend on the different parameters. Our paratactical goal was to obtain an efficient cost function rather than the optimal one.

An efficient cost function needs to have a global minimum for the targeted set of parameters. Verifying this condition requires that the expected parameter values are known. Consequently, the selection of metrics was performed on the numerical calibration phantom for which $\chi_c$ is controlled.

Metrics based on geometric criteria (angles between the threads, straightness of the threads) and metrics of image quality (image amplitude, sharpness) on a local scale around each thread and on the entire image were tested for the PA image and the US image. Metrics describing the PA and US image superposition (image cross-correlation, distance between the reconstructed threads) were also considered. The initial screening of the metrics implied graphical visualization of the metric variations with the parameter values. To facilitate this visualization, only one parameter of $\chi_c$ was varied while all the other parameters were held constant and equal to their expected values. In this first stage, we chose to set all the parameters of $\chi_c$ to zero. Then, the angular parameters were varied over a range of [−5°;5°] by steps of 0.5°, the distances over a range of [−3 mm; 3 mm] by steps of 0.3 mm, and the time parameter over a range of [−0.5μs; 0.5μs] by steps of 0.1 μs.

We found that the cost function could not be defined with a single metric because none of the tested metrics had a global extremum for all the parameters. Therefore, we performed a careful selection of metrics that were subsequently combined to obtain the cost function. Because of the diversity of metrics, we chose to give a score to each of them and for each parameter. Criteria for the score were the presence of one global extremum at the expected value and steep variations. A selection among the tested metrics was then performed keeping the ones with highest scores while making sure that all the parameters were covered by at least one metric.

The selected metrics are presented below (II.C.3.c). We combined them with a product. Compared to a weighted sum, the combination with a product enables mixing metrics with different scales without the need to perform a prior normalization, but may be more sensitive to fluctuations (noise). With the same procedure as for the test of each metric, we tested that the defined cost function had a minimum at the expected value when each parameter was varied individually (all the other parameters remained constant). For this validation, the set of parameters was fixed to non-zeros values to avoid a bias.

*c)* *Selected metrics for the cost function*

In this section, we present the image-based metrics selected to build the cost function. We determined that some metrics were more sensitive to parameter variations for Threads 2 and 4. Thereby, several quantities were computed for these two threads. First, a linear regression was performed with the five centroids (one per slice) of each thread and the coefficient of determination was computed. The mean coefficient of determination over the two threads gave $R^2_{US}$ and $R^2_{PA}$ for the US images and the PA images, respectively. $R^2_{US}$ and $R^2_{PA}$ equals one when the threads are reconstructed as straight structures in the volume and when the image quality allows adequate segmentation of the threads. $R^2_{US}$ and $R^2_{PA}$ were found to be sensitive to errors for the parameters Pitch, $\Delta x$ and $\Delta z$. Second, the mean distance of $d_{PA-US}$, named $D_{PA-US}$, was computed over the two threads and the five slices. $D_{PA-US}$ evaluated the superposition of the PA and US images. $D_{PA-US}$ is expected to be equal to zero for the correct set of parameters. $D_{PA-US}$ was found to be sensitive to errors in the parameter $t_{0\ US}$. Additionally, a thresholding of $D_{PA-US}$ was found efficient to reject sets of parameters leading to a degraded line spread function either in US or in PA. $D_{PA-US}$ is expressed in mm and the cost was set equal to zero for $D_{PA-US} > 1$ mm (meaning that the set $\chi$ is rejected). Finally, the local normalized variance of the US images was computed in a square region of 2 mm-width (twice the translation step) around the centroid. The normalized variance quantifies variations in the pixel values about the mean. It is equal to the variance of the pixel values over their mean. This measurement of the image sharpness was reported for an autofocus method [33]. The mean of the local normalized variance over the two threads and over the five slices was named $N_{V\ US}$. $N_{V\ US}$ is expected to be maximal for the correct set of parameters. $N_{V\ US}$ was found to be sensitive to errors in the parameters: Pitch, Yaw, $\Delta x$, $\Delta z$, $\theta$.

Finally, a normalized two-dimensional squared gradient of the entire US images (four threads) was computed as a sharpness metric:

$$S_{N\ 2D} = \frac{1}{\sum_{x,z} f(x,z)} \left( \sum_{x,z} (f(x+1,z) - f(x,z))^2 + \sum_{x,z} (f(x,z+1) - f(x,z))^2 \right) \quad (7)$$

where $f(x,z)$ represents the gray level intensity in the 2D ultrasound image. The mean of the normalized squared gradient over the five slices was named $S_{N\ US}$. $S_{N\ US}$ was sensitive to errors in the parameters: Roll, Pitch, Yaw, $\Delta x$, $\Delta z$ and $t_{0\ US}$.

The selection of the metrics $N_{V\ US}$ and $S_{N\ US}$ highlight that the normalized variance and normalized squared gradient were more efficient for US images than for PA images. This could be explained by the fact that the US image reconstruction combines data for five steered angles, which may induce stronger variations when the parameters are away from their expected values.

We combined $R^2_{US}$, $R^2_{PA}$, $D_{PA-US}$, $N_{V\ US}$ and $S_{N\ US}$ with a product and the cost was then defined by:

$$Cost = \begin{cases} 0 & \text{for } D_{PA-US} \geq 1\ \text{mm} \\ -R^2_{US} \cdot R^2_{PA} \cdot (1 - D_{PA-US})^2 \cdot N_{V\ US} \cdot S_{N\ US} \end{cases} \quad (8)$$

*4) Calibration algorithm: Optimization algorithm*

The calibration algorithm relies on the cost function. An initial combination of 7 parameters was given as an input. Two steps were then applied. First, several combinations of parameters were proposed, in which each parameter was drawn at random following a normal distribution around the initial guess. Costs of these combinations were computed until reaching a total of 100 combinations with non-zero cost. The combination with the smallest cost was used for the second step: the application of a Particle Swarm optimization algorithm. This heuristic algorithm was chosen over a convergent iterative method because of the global optimization problem at hand.

With the numerical calibration phantom, we found a variability of the output combination and of the associated cost when the calibration algorithm was repeated on the same dataset. This indicates local minima of the cost function. To mitigate this variability, the optimization algorithm was repeated 20 times on the same dataset. The 20 combinations were then sorted by increasing cost and the final combination was obtain by calculating the median of each parameter on the combinations with the 5 lowest costs. The median was chosen for its ability to limit the weight of strong outliers on the solution.

*5) Metrics for the variability*

*a)* *Acceptability range*

The variability of the determination of the parameters was compared to an acceptability range. Three acceptability ranges were defined depending on the unit of the parameter. The acceptability range for length parameters ($\Delta x$, $\Delta z$) was set equal to the wavelength $\Lambda_c$. For $t_{0\ US}$, we used the wave period at the central frequency of the transducer: 0.2 μs. For the angles, we considered an axial deviation of $\Lambda_c$ seen from the lateral aperture of the array as a significant error. As the interelement spacing of the array and the element size can be considered equal to $\Lambda_c$, the lateral aperture dimension is equal to $64 \cdot \Lambda_c$ and the acceptability range for angles was then set to $\sin^{-1}\frac{\Lambda_c}{64 \cdot \Lambda_c} = 15.6$ mrad (0.9°). This value also corresponds to the angular resolution of the array in the lateral direction [36].

*b)* *Variability quantifications*

To assess the variability induced by the optimization algorithm, the numerical calibration phantom was used. The simulated set of parameters $\chi_c$ reflected the experimental data (mean over 10 experiments). Absolute difference between obtained and expected ($\chi_c$) parameters were computed and divided by the acceptability threshold to be expressed as a percentage. This metric evaluated the accuracy of the calibration method and therefore is named Ac.

To assess inter-acquisition variability on experimental data, parameters were obtained for 10 acquisitions. These parameters were compared to the mean parameter over the 10 acquisitions. For each parameter, the mean absolute difference of the obtained outputs compared to the mean value was computed (also named mean absolute deviation) and divided by the acceptability threshold to be expressed as a percentage. This parameter assesses the repeatability of the entire calibration process and is called Rp.

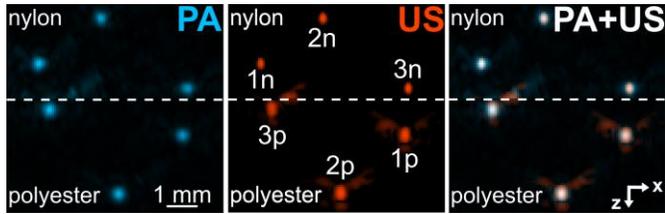

Fig. 3. Image reconstruction of the central plane of Ph1 (y=0). All images are 7-mm wide (x-axis) by 6.7-mm high (z-axis) and are centered at (x=2 mm, z=25mm). The horizontal white dashed line represents the separation between nylon and polyester zones. Threads with the same number are symmetrical and n stands for nylon and p for polyester. PA image (left) is presented on a linear scale while the US image (center) is presented in dB with a threshold at -30dB.

6) *Metrics for the spatial resolution and superposition*

We quantified the superposition of the US and PA images using a second phantom comprised of threads (see Ph1, in section II.D) with a dual contrast. The quantification method assesses the distance between the images of each thread in USI and in PAI. As for the image processing used in the calculation of $D_{US-PA}$ on the calibration phantom, we determined the positions of the centroids (both in the US images and the PA images) for each thread and 5 slices located at y = i×3 mm with $i \in [\![-2; 2]\!]$. Each slice was reconstructed with pixel sizes py × pz = 71 µm × 71 µm. The mean distance of $d_{US-PA}$ (see equation (6)) and its standard deviation over the slices and over the threads of the same material (nylon or polyester) measured the superposition quality.

For each nylon thread, the spatial resolution was estimated by fitting the images in each slice *i* with a 2D-Gaussian model equation:

$$f(x,z) = A \cdot \exp\left(-\left(\frac{(x-x_0)^2}{2\sigma_x^2} + \frac{(z-z_0)^2}{2\sigma_z^2}\right)\right) \quad (9)$$

where $A$ is the amplitude, $x_0, z_0$ are the centroid positions and $\sigma_x, \sigma_z$ are the standard deviations along $x$ and along $z$. The full width half maximum was calculated along $x$ and $z$ as $FWHM_i = 2\sqrt{2 \ln 2} \times \sigma_i$. Because nylon threads can be considered small compared to $\Lambda_c$, the FWHM is an estimate of the width of the Line Spread Function (LSF).

D. *Imaging phantoms*

The first phantom (Ph1) was a wire phantom designed to evaluate the co-registration capabilities. Ph1 differs from the calibration phantom in the spatial arrangement of the threads and because half of the threads are of a different material. Three 20-µm diameter black nylon threads (NYL02DS, Vetsuture, France) and three black polyester threads (Coat Epic 150) were mounted on a 3D-printed frame similar to Fig. 2(b). They were arranged symmetrically with respect to the center of the frame and with various angles.

The second phantom (Ph2) was made by suspending 2% w/v agar powder (A1296, Sigma Aldrich, St. Louis, MO, USA) and 1% w/v cellulose powder (Sigmacell cellulose Type 20, Sigma Aldrich) in water to form a gel cast in a cylindrical mold (20-mm diameter) with three mold-length, cylindrical solid inclusions that were 5 mm in diameter. Within this gel, the cellulose forms particles of approximately 20 µm which act as US scatterers to mimic the US scattering properties of biological tissues. The agar-cellulose solution was heated to 85°C and poured into the mold. When the mold was half full, 100-µm-diameter black polyethylene microspheres (BKPMS 90–106 um, Cospheric, Santa, Barbara, CA, USA) were spread on the superior interface and were trapped at the interface during the solidification of the gel. These black microspheres act as optical absorbers that can generate a PA signal. The mold was then filled with the hot agar solution. When the gel solidified, the cylindrical inclusions were removed and filled with water. The microspheres remained embedded in the gel. Ph2 was placed so that the cylindrical holes and the plane of spheres were parallel and perpendicular to the rotation axis, respectively.

TABLE II
RESOLUTION AND SUPERPOSITION ASSESSMENT

| | FWHM nylon (µm) | | Superposition (µm) | |
|---|---|---|---|---|
| | US | PA | nylon | polyester |
| along x (mean ± std) | 212 ± 18 | 387 ± 30 | 15 ± 9 | 23 ± 17 |
| along z (mean ± std) | 296 ± 24 | 379 ± 21 | | |

The third phantom (Ph3) was prepared with agar powder (2% w/v) and cellulose powder (1% w/v) in water for the first half and with agar powder (2% w/v) for the second half. Two crossed 20-µm diameter black nylon threads were embedded in the gel.

III. RESULTS

A. *Accuracy and repeatability of the calibration*

The accuracy and the repeatability of the calibration outputs are presented in Table. I, in percentage of the acceptability range. First, it can be noticed that all the 7 parameters are fully within the acceptability ranges, for both Ac and Rp.

For the accuracy, the calibration outputs were compared to the ground truth thanks to the numerical simulation. The mean Ac over the 7 parameters was found to be equal to 26% and Ac had a maximum of 55% for the Roll parameter. We can thus consider that the developed calibration method enables to accurately determine the set of parameters.

To evaluate the repeatability, ten acquisitions were performed on various days (distributed in three imaging sessions over one week). For each acquisition, the optical fibers and the phantom were repositioned to avoid any bias. Rp estimates the mean absolute deviation of each parameter. The mean Rp over the 7 parameters is around 16% and a maximum of 39% was reached for Δz. Therefore, the calibration is repeatable and stable over time.

Despite slight variations in the evaluation of each parameters, the accuracy and the repeatability of the calibration method are highly satisfying. We can therefore consider the developed calibration method to be reliable and robust.

B. *Superimposition quality and spatial resolution*

Fig. 3 presents PA/US images of the slice y =0 (center of the linear array) of Ph1. Ph1 aims to test if the experimental calibration remains valid for a phantom different than the calibration phantom in the spatial arrangement of the threads, but also in the thread material. The phantom Ph1 is comprised of three nylon threads and three polyester threads arranged so that each thread has a different orientation. Only the threads numbered 2 were set parallel to the **y**-axis. For an easier comparison between the two materials, Ph1 was built so that each of the three nylon threads had a symmetrical polyester thread with respect to the center of the frame. In Fig. 3, symmetrical threads have the same number and the suffixes 'n' and 'p' refer to nylon and polyester threads, respectively. For instance, the thread 1n is symmetrical to 1p, with respect to the center of the image. The nylon threads were grouped on the top part of the phantom while the polyester threads are grouped on the bottom part. Video 2 displays rotating MAP images around z of Ph1 and therefore shows the spatial arrangement of the threads. As nylon threads are thinner than polyester ones, both US and PA signals were weaker for nylon. The amplitude in the reconstructed image was 3 times smaller in PA. To facilitate the visualization in Fig. 3, the upper and the lower part of the volumetric images were normalized by their local maximum and not the global one. The separation between the two parts is illustrated by the horizontal dashed white line in Fig. 3. We can visually see that polyester threads appear larger than the nylon ones both on the PA and US images. This is expected given their larger diameter. However, the superposition of the images can be observed for the two materials and for all the threads regardless of their orientation or position in space (Fig. 3 and Video 2). The blue lateral halo around the white spots in the combined image (Fig. 3 left) indicates that the lateral resolution is wider for PA images.

For a more quantitative description, Table. II presents the superposition distance and the FWHM calculations. The superposition distance between the center of a thread in PA and in US is small compared to the US wavelength $\Lambda_c$ (less than 10% of $\Lambda_c$) and smaller than the spot obtained for each object in the image. The superposition distance is not significantly different for the two materials. This result indicates that the calibration enables the superposition of nylon thread with different orientations than in the calibration phantom, and it validates that the calibration ensures the superposition for objects with a dual contrast but in another material. The standard deviation computed over the three threads and five imaging planes is also small compared to $\Lambda_c$. This result shows the low dispersion of the superimposition distance both with the thread orientation and with the spatial position in the imaged volume.

The FWHMs calculated for the nylon threads provides an estimate of the LSF because the thread diameter is much smaller than $\Lambda_c$. We first notice that for both US and PA and for both x and z directions, the FWHM is on the order of magnitude of $\Lambda_c$. Along the x-direction, the resolution is limited by the diffraction. The high resolution in the x-direction results from the large angular aperture provided by the rotation scan and the synthetic aperture approach. For comparison, the $FWHM_x$ was on the order of 1 mm [27] for a translation-only scan with the same US array. The low standard deviation of $FWHM_x$ indicates that the resolution is independent of the position of the object in the volume and its orientation. The spatial homogeneity associated with the rotate-translation scheme and previously observed independently in USI [27] and PAI [26] is then confirmed for the simultaneously co-registered imaging.

In each direction, the US resolution is slightly better than the PA resolution (70-100% of $\Lambda_c$ for US vs 130% of $\Lambda_c$ for PA). Along x, two main factors can explain the $FWHM_x$ differences between US and PA. First, PA images rely on US signals produced by the illuminated object and not on backscattered US signals generated by the US array. For small objects, the US frequency spectrum recorded by the array is then usually broader in PAI than in USI, and especially contains low frequencies which may decrease the diffraction-limited resolution. Second, for one laser excitation, five tilted plane waves are emitted, which increases the number of independent views and the spatial sampling in USI compared to PAI. This sampling factor has been shown to influence the resolution along x [27]. Along z, which corresponds to the axial direction, the LSF is mainly influenced by the pulse duration. The pulse-echo mechanism can explain the better resolution of the US images compared to PA.

### C. Complementary distributions of US and PA contrasts

To further demonstrate the advantages of the dual modality imaging, phantoms with complementary contrasts were designed and produced. The images of Ph2 and Ph3 are displayed in Fig. 4 and 5, respectively.

For the ultrasound contrast, Ph2 is a homogeneously scattering medium (agar with cellulose) with three cylindrical and anechoic holes filled with water. For the PA contrast, numerous black-dyed micro-spheres are arranged in a central plane (Fig. 4). The optical absorption of the agar gel and the water are negligible. In the first row of the Fig. 4, a MAP image along y is shown for PAI, USI, and the superposition of both. Rotating MAP images round the z-axis are presented in Video 3. The 20-mm diameter cylindrical shape of the phantom could be retrieved both in PAI and USI. One can note the homogeneous image quality in the xz-plane for both modalities. In the second row of Fig. 4, images of a slice perpendicular to the z-axis are presented. The slice was chosen to cut two of the three holes. The homogeneity of the USI image along the y-axis can be observed. As expected, the three holes appear with a negative contrast for both PAI and USI independently of their spatial position. The holes allow to further validate the superimposition of the two modalities. They are concentric in USI and in PAI. We can notice that the outlines of the holes are blurrier and the diameter of holes seems smaller on the US images. This effect can be attributed to the stronger side lobes in the US images induced the log compression (display in dB) of the color scale compared to the linear scale used in PAI. Such side lobes are also visible in Fig. 2(c). In US imaging, the microspheres have a contrast relative to the agar matrix only when they are numerous and clumped (diagonal line at the bottom of Fig. 4-second row), or when they are distributed along a surface that visually integrates their contribution in MAP rendering (Video 3). However, in Fig. 4-second row, microspheres are hardly visible in the middle and top parts of the US image, while they appear with a strong contrast in PA imaging. With the superposition, both the agar and the microspheres are visible. This is of interest to assess the distribution of PA contrast agents in an organ which is homogenously echoic for instance. The US images gives the contour of the phantom, which is similar to the anatomical context, while the PA image gives the distribution of the marked spheres which are analogous to a contrast agent. In a clinical application, the PA contrast agents could be therapeutic nanoagents accumulating locally. Holes are mimicking bodies with a negative contrast such as cysts.

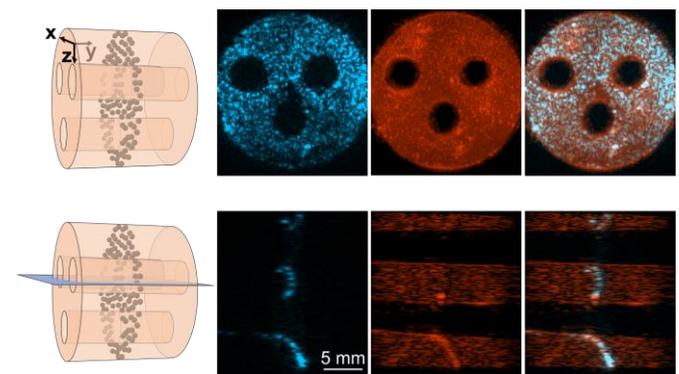

Fig. 4. Image reconstruction of Ph2. Schematic drawing of Ph2 are displayed in the left column. The first row displays MAP images along y while the second row shows a slice perpendicular to the z-axis. PA images are presented in linear color scale. However, because a few bright spots (probably clusters of microspheres) were dominating the color scale and hiding the rest, voxel values were saturated at 40% of their maximum before normalization. US images were thresholded at -40 dB. Fist row images are 19-mm wide (x-axis) by 21-mm high (z-axis) and centered at (x=-0.6 mm, z=24.3 mm); second row images are 19-mm wide (x-axis) by 21-mm high (y-axis) and centered at (x=0 mm, y=0 mm).

For the second phantom Ph 3, media with two different ultrasound contrasts were used: agar with cellulose is more echogenic that agar alone. Two black nylon threads bring the optical absorption contrast. Images of Ph3 are presented in Fig. 5. The US image in the first row (MAP along z) and the rotating MAP images in Video 4 clearly show the contrast between the two blocks of gel. The nylon threads are visible in the MAP US images, but mainly due to their elongated shape as they are barely differentiable from the surrounding speckle in the slice perpendicular to the y-axis and intersecting the two threads in the agar with cellulose part (second row of Fig. 5). Agar and agar with cellulose do not have any contrast in PA. However, the threads show a homogenous and high contrast in PAI over the two agar blocks. This phantom could typically mimic a blood vessel (black thread), highly visible in PAI, perfusing two different organs with different echogenicity.

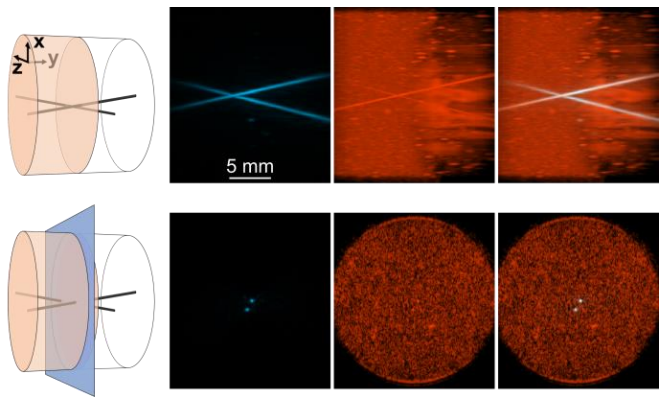

Fig. 5. Image reconstruction of Ph3. Schematic drawing of Ph3 are displayed in the left column. The first row displays MAP images along z that are 19 mm wide (y-axis) by 21 mm high (x-axis) while the second row shows a slice perpendicular to the y-axis that is 19 mm wide (x-axis) by 21-mm high (z-axis) and centered at x=0 mm, z = 25.4 mm. US images were thresholded at -40 dB. Along y-axis, left part is agar with cellulose and right part is agar alone.

## IV. DISCUSSION

We demonstrated high-quality, volumetric and simultaneously co-registered PAI and USI. The simultaneous dual imaging was made possible by the use of a linear US transducer array, and the high quality (resolution, contrast, visibility) of the volumetric images resulted from the large synthetic angular aperture enabled by the rotate-translate scan geometry. Images showed a homogenous quality over a large imaged volume (cylinder of diameter of 21 mm and length of 19 mm). The effective synthetic aperture for both imaging modalities and the superposition of the PA and US images for features having a dual contrast required an accurate determination of the positions of the US array. To this end, we developed and validated a calibration method, which was determined to be both accurate and repeatable. This initial calibration process allowed the reconstruction of images from subsequent acquisitions without the need of fiducial markers on every imaged volume under conditions for which the speed of sound can be considered constant in the imaged volume. We demonstrated the superposition of PA and US images with phantoms having a dual contrast and the complementarity of the mapped information with phantoms having complementary spatial distributions of US and PA contrast agents.

The calibration was based on the combination of a dedicated calibration phantom, a cost function and an optimization process. Seven parameters were determined for our scan geometry to obtain superimposed and sharp PA and US images. The mechanical mount and the position sensors of the stages ensure that the determined parameters remain valid for subsequent scans as long as no deformation or accidental misalignment of the array occurs. For this study, no significant loss of calibration was observed over a period of one week. The calibration phantom was easy to build and was comprised of four well-separated threads to facilitate the identification and measurement of local image properties. The cost function used only five imaging planes to avoid the reconstruction of the entire volume and the associated long computation time. Because metrics with the sharpest variations with regards to the parameters were selected for the cost function, it was dominated by the properties of the US image but was also influenced by the properties of PA images. For the optimization algorithm, we found that the selection of the initial guess was a crucial step due to the presence of local minima in the cost function. Additionally, the particle swarm optimization was found to be more effective to determine a solution close to the ground truth than a downhill simplex method. The refinement of the optimization algorithm is beyond the scope of this study since we focused on obtaining and effective method that is simple to apply. Further refinement will be considered in a future study.

This calibration procedure was found to be effective for our system although it requires the initial imaging of a dedicated phantom. In the past decade, several multiperspective or multiview imaging systems with US transducer arrays have been developed both in PA imaging and in US imaging, independently, leading to various calibration procedures. For US imaging, coherent compounding of images acquired with two US transducer arrays have been developed in 2D [37] and in 3D [38]. Dedicated calibration phantoms were also used that consisted in isotropic scatterers (5 wires in 2D and 3 spheres in 3D). The phantoms were combined with a cost function linked to the coherence of the echoes from the isotropic scatterers received by the different elements of the array to determine geometrical parameters (4 parameters in 2D and 6 in 3D). A simplex search method was used. The calibration showed an improvement in the contrast and in the resolution of the images but the use of isotropic scatterers resulted in long computation time of volumetric images during the calibration process. Incoherent compounding of US images was also investigated for 2D [39] and 3D [40] imaging. The spatial transformation matrix between the positions of the arrays were assessed directly using the images acquired on the sample of interest. This approach could enable free hand scanning and imaging of moving organs but could not achieve a synthetic aperture approach which, therefore, limits the final image quality. In PA imaging, multiperspective or multiview imaging is almost inherent to the modality since PA tomography with spherical and hemispherical scans of an US detector has been investigated in the early PA scanners [41]. However, planar detection geometries able to acquire 3D PA images have been angulated to enhance the image view with a synthetic aperture approach [42], [43]. In ref [42], two planar arrays were assembled in a rigid configuration. An initial calibration was performed by imaging a dedicated phantom comprised of three threads with different orientations with each array independently. Thread position and orientation were then identified in each segmented volume to determine the rigid transformation between arrays. In ref [43], fiducial markers were incorporated in the imaged region. Multiview imaging was also performed by stitching together volumetric images [44], which is similar to incoherent compounding and relies on the features obtained in partially overlapping images. In brief, no standard calibration exists either in US or PA imaging. To date, only dedicated calibration phantoms have been shown to allow synthetic aperture reconstruction without fiducial markers. Additionally, when using a single imaging modality, some reconstruction parameters may compensate for errors in other reconstruction parameters. For volumetric and simultaneous PA and US imaging, the superposition and sharpness of the dual-mode tomography can only be obtained if each parameter is determined accurately. In particular, this accuracy is necessary because of the propagation times between transducer elements and the voxels is different for PA imaging and US imaging.

Co-registered volumetric PA and US maps provide complementary information (Fig. 4 and 5). In the framework of longitudinal studies with repeated imaging sessions, the US image is expected to give the anatomical reference so that the PA response can be better positioned within the anatomy and so that robust co-registration can be performed between data acquired at different time points. PA response can be applied to reveal molecular or functional phenomena with a slow kinetic, such as the accumulation of a nanoparticular contrast agents with a long circulation time.

For other simultaneously co-registered multimodal imaging such as PET/CT, it has been shown that the anatomical imaging modality contains information that can be used to improve the image quality of the functional imaging modality. Recent work on the reconstruction of 2D PA images has shown that the image quality of blood vessels in PA images can be improved using structural information from ultrasound images [45]. Additionally, light fluence distribution could be modeled using US images to further improve PA image quality [46].

Our image reconstruction algorithms adapted conventional delay-and-sum beamforming approaches to the tomographic system. While these algorithms led to high quality images for the tested phantoms, more complex image reconstruction algorithms will be investigated in futures studies to incorporate acoustic properties of the sample or of the transducer. Indeed, heterogeneous speed of sound or aberrating layers in the sample would result in degraded image quality with our algorithm but may be considered with iterative reconstruction approaches [47], [48]. Modeling the spatial impulse response of the transducer in the reconstruction algorithm could allow a further improvement of the image quality and a reduction of the number of scan positions and, consequently, of the scan duration [49]. Motion of the sample during the scan is expected to induce blurring artifacts. However, because of the interlaced PA and US events, motion artifacts are expected to be similar for the two imaging modalities. These artifacts may be reduced by gating the acquisition with physiological signals to select phases for which the imaged volume is still or has returned to the same position.

The study presented here was performed at a single optical wavelength. Multispectral approaches [50] will be investigated with the developed scanner to enable discrimination between different chromophores and contrast agents and to evaluate physiological parameters such as the oxygen saturation. On the US side, the system presented here operated at an US center frequency of 5 MHz. However, the rotate-translate scan can be scaled to other ultrasound frequencies to gain sensibility and information on other spatial scales [26], [51]. The spatial resolution increases with the US frequency of the array for both PA and US imaging, but higher frequencies will also improve the sensitivity to small absorbing regions in PA imaging. Linear US arrays are commercially available in a wide range of center frequencies.

Building on the *in vitro* proof of concept presented here, we are currently adapting the scanner for *in vivo* imaging with the addition of an acoustic coupling system to remove the large water tank. A small water tank with an acoustically transparent membrane sealing the bottom [20] will be investigated. For the translation toward clinical applications, the bulky motorized stages will be replaced by a scanning system with a smaller foot print such as those that have been previously developed for translational scans [20], [21].

The imaged volume (cylinder of diameter of 21 mm and length of 19 mm centered at a depth of 25 mm), sub-millimeter spatial resolution and volumetric imaging rate (~20 s per volume) can usefully offer dual-mode contrast for pre-clinical investigations of murine models [44], [52]. Potential clinical applications that could benefit from the combined, volumetric PA and US imaging offered by this technique include the evaluation of inflammation (finger and hand joints [53] for instance) and cancerous lesions (such as thyroid cancer [54]) in relatively superficial zones of the body.